\documentclass[pra,twocolumn,nofootinbib,floatfix]{revtex4}

\usepackage[usenames]{color}
\usepackage{amsmath}
\usepackage{amssymb}
\usepackage{float}
\usepackage{graphicx}
\usepackage{xspace}
\long\def\symbolfootnote[#1]#2{\begingroup%
\def\thefootnote{\fnsymbol{footnote}}\footnotetext[#1]{#2}\endgroup}

\begin{document}

\title{Superfluid behaviour of a two-dimensional Bose gas}

\author{R\'{e}mi Desbuquois$^{1}$}
\author{Lauriane Chomaz$^{1}$}
\author{Tarik Yefsah$^{1\dag}$}
\author{Julian L\'{e}onard$^{1\ddag}$}
\author{J\'{e}r\^{o}me Beugnon$^{1}$}
\author{Christof Weitenberg$^{1*}$}
\author{Jean Dalibard$^{1}$}

\date{21 May 2012}

\affiliation{
   $^1$Laboratoire Kastler Brossel, CNRS, UPMC, Ecole Normale Sup\'erieure, 24 rue Lhomond, F-75005 Paris, France
}

\begin{abstract}
Two-dimensional (2D)  systems play a special role in many-body physics. Because of thermal fluctuations, they cannot undergo a conventional phase transition associated to the breaking of a continuous symmetry \cite{Mermin:1966}. Nevertheless they may exhibit a phase transition to a state with quasi-long range order  via the Berezinskii-Kosterlitz-Thouless (BKT) mechanism \cite{Minnhagen:1987}. A paradigm example is the 2D Bose fluid, such as a liquid helium film \cite{Bishop:1978}, which cannot Bose-condense at non-zero temperature although it becomes superfluid above a critical phase space density. Ultracold atomic gases constitute versatile systems in which the 2D quasi-long range coherence and the microscopic nature of the BKT transition were recently explored \cite{Hadzibabic:2006,Clade:2009,Tung:2010}. However, a direct observation of superfluidity in terms of frictionless flow is still missing for these systems. Here we probe the superfluidity of a 2D trapped Bose gas with a moving obstacle formed by a micron-sized laser beam. We find a dramatic variation of the response of the fluid, depending on its degree of degeneracy at the obstacle location. In particular we do not observe any significant heating in the central, highly degenerate region if the velocity of the obstacle is below a critical value.
\end{abstract}

\maketitle

\symbolfootnote[2]{present address: Department of Physics, MIT-Harvard Center for Ultracold Atoms, and Research Laboratory of Electronics, Massachusetts Institute of Technology, Cambridge, Massachusetts 02139, USA.}
\symbolfootnote[3]{present address: Institute for Quantum Electronics, ETH Z\"urich, 8093 Z\"urich, Switzerland.}
\symbolfootnote[1]{Electronic address: \url{christof.weitenberg@lkb.ens.fr}}

`Flow without friction' is a hallmark of superfluidity \cite{Leggett:2006}. It corresponds to a metastable state in which the fluid has a non-zero relative velocity $v$ with respect to an external  body such as the wall of the container or an impurity. This metastable state is separated from the equilibrium state of the system ($v=0$) by a large energy  barrier, so that the flow can persist for a macroscopic time. The height of the barrier decreases as $v$ increases, and eventually passes below a threshold (proportional to the thermal energy) for a critical velocity $v_c$. The microscopic mechanism limiting the barrier height depends on the nature of the defect and is associated to the creation of phonons and/or vortices \cite{Leggett:2006}. While the quantitative comparison between experiments and theory is complicated for liquid $^4$He, cold atomic gases in the weakly interacting regime are well suited for precise tests of many-body physics. In particular, superfluidity was observed in 3D atomic gases by stirring a laser beam or an optical lattice through bosonic \cite{Raman:1999,Onofrio:2000,Raman:2001,Engels:2007,Neely:2010} or fermionic \cite{Miller:2007} fluids and by observing the resulting heating or excitations. Here we transpose this search for dissipation-less motion to a disc-shaped, non homogeneous 2D Bose gas. We use a small obstacle to locally perturb the system. The obstacle moves at constant velocity on a circle centred on the cloud, allowing us to probe the gas at a fixed density. We repeat the experiment for various atom numbers, temperatures and stirring radii and identify a critical point for superfluid behaviour.

\begin{figure}
	\begin{center}
		\includegraphics[width=1.0\columnwidth]{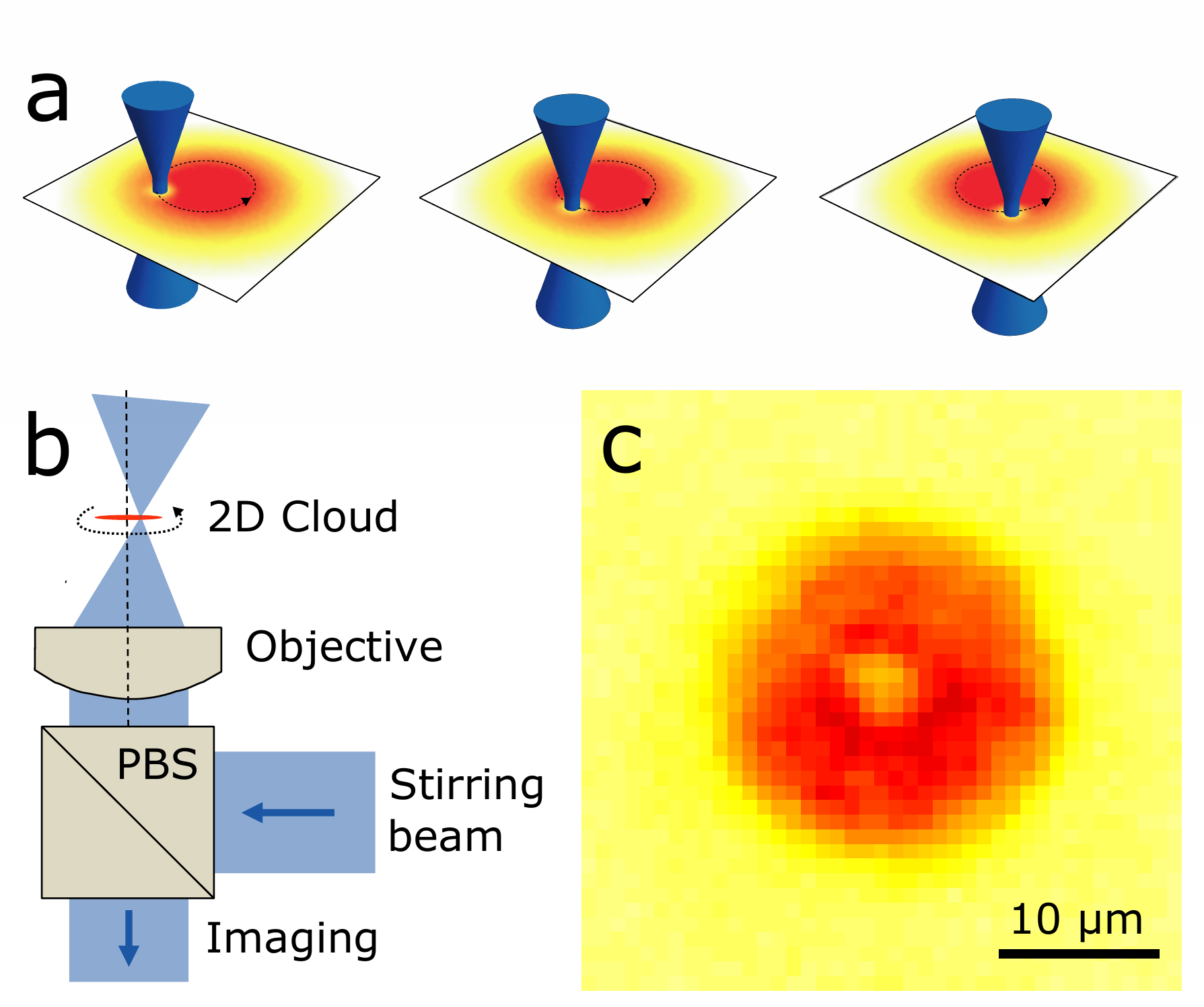}
	\end{center}
\caption{{\bf Stirring a 2D Bose gas.} {\bf a}, A trapped 2D gas of $^{87}$Rb atoms is perturbed by a focussed laser beam, which moves at constant velocity on a circle centred on the cloud. The stirring beam has a frequency larger than the $^{87}$Rb resonance frequency (`blue detuning' of $\approx\,2\,$nm) and thus creates a repulsive potential which causes a dip in the density profile.
{\bf b}, The stirring beam is focussed onto the 2D cloud via a microscope objective of numerical aperture 0.45, which is also used for imaging. We overlap the two beam paths with a polarizing beam splitter cube (PBS). The position of the stirring beam is controlled by a two-axis piezo-driven mirror.
{\bf c}, \emph{in situ} false-color image of the 2D cloud in the presence of the laser beam (average over six images). From the dip in the density we deduce the waist of the laser beam as $w_0 = 2.0(5)\,\mu$m. In this image, the intensity of the beam is chosen three times higher than in the stirring experiment to make the hole well visible even in the center of the cloud.
We use similar images, but with the stirring beam switched off, to determine the temperature $T$ and the chemical potential $\mu$ from a fit of the Hartree-Fock prediction to the wings of the cloud \cite{Yefsah:2011}.
\label{fig:scheme}}
\end{figure}

\begin{figure*}[!t]
	\begin{center}
		\includegraphics*[width=1.8\columnwidth]{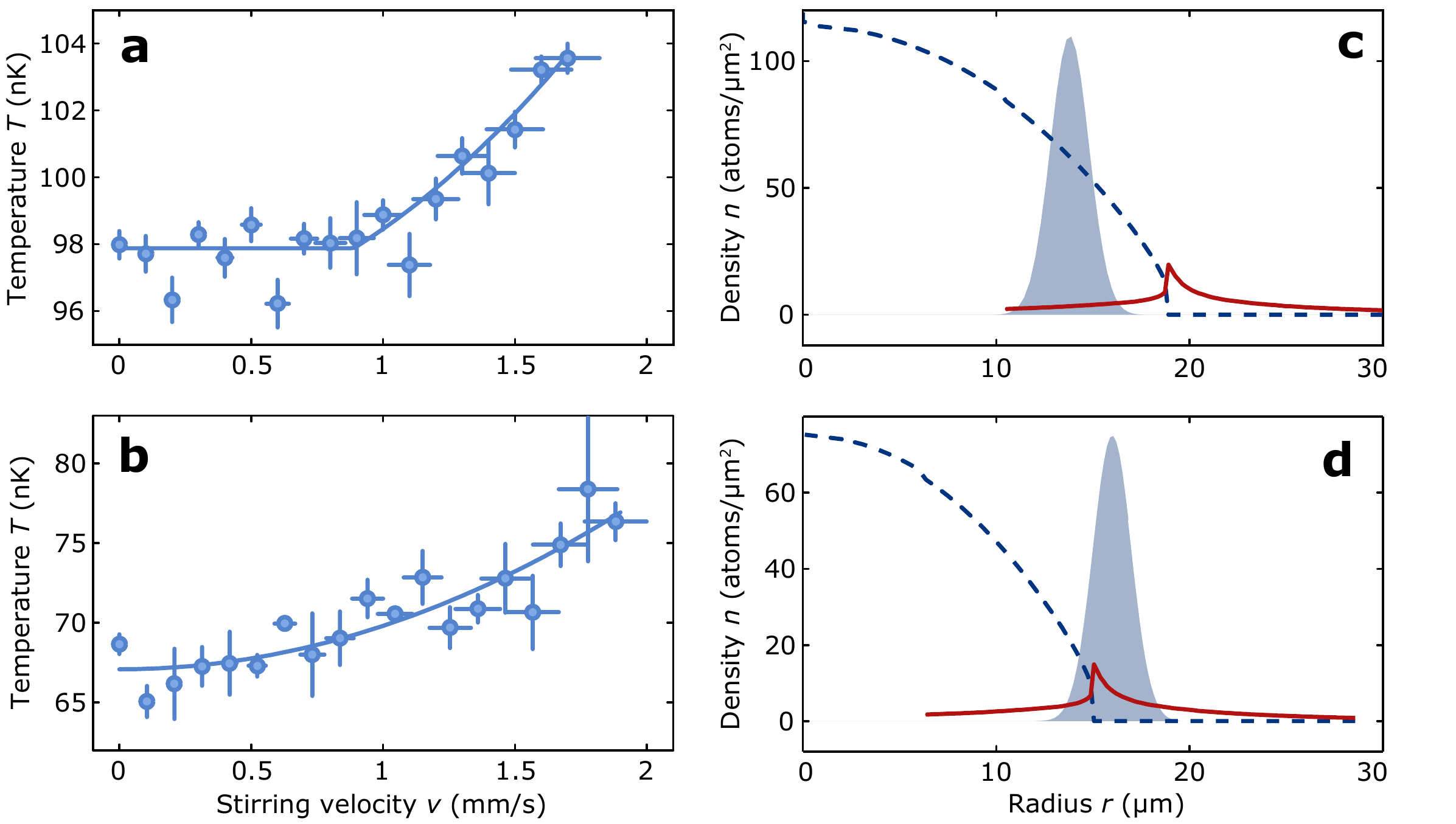}
	\end{center}
\vspace{-0.3cm}
\caption{{\bf Evidence for a critical velocity}. Two typical curves of the temperature after stirring the laser beam at varying velocities. {\bf a}, In the superfluid regime, we observe a critical velocity (here $v_c=0.87(9)\,$mm/s), below which there is no dissipation. {\bf b}, In the normal regime, the heating is quadratic in the velocity.
The experimental parameters are ($N$, $T$, $\mu$, $r$)=($87000$, $89\,$nK, $k_{\rm B}\times 59\,$nK, $14.4\,{\rm \mu m}$) and ($38000$, $67\,$nK, $k_{\rm B}\times 39\,$nK, $16.6\,{\rm \mu m}$)  for {\bf a} and {\bf b}, respectively, yielding $\mu_{\rm loc}/k_{\rm B}T=0.36$ and $\mu_{\rm loc}/k_{\rm B}T=0.04$.
The data points are the average of typically ten shots. The $y$ error bars show the standard deviation. The $x$ error bar denotes the spread of velocities along the size of the stirring beam ($1/\sqrt{e}$ radius).
The solid line is a fit to the data according to equation\,(\ref{eq:fitFunction}). Note that the three low-lying data points in {\bf a} correspond to the completion of half a turn and are correlated to a displacement of the cloud, which may be responsible for the observed 1.5\,nK temperature shift of these points.
{\bf c} and {\bf d}, Calculated radial density distribution for the clouds in {\bf a} and {\bf b}, respectively. The dashed blue curve shows the superfluid density, the solid red curve shows the normal density. The stirring beam potential is indicated by the grey shaded area (in arbitrary units). The densities are calculated via the local density approximation from the prediction for an infinite uniform system \cite{Prokofev:2002}. The jump of the superfluid density from zero to a universal value of $4/\lambda_{\rm dB}^2$ (where $\lambda_{\rm dB}$ is the thermal de Broglie wavelength) is a prominent feature of the BKT transition. The normal density makes a corresponding jump to keep the total density continuous.
\label{fig:heating}}\vspace{-0.3cm}
\end{figure*}

Our experiments are performed with 2D Bose gases of $N=35000$ to 95000 $^{87}$Rb atoms confined in a cylindrically symmetric harmonic potential $V(r)+W(z)$ \cite{Yefsah:2011}. The trap frequencies are $\omega_{r}/2\pi=25.0(5)\,$Hz in the horizontal plane and $\omega_z/2\pi=1.4\,(1)\,$kHz in the vertical direction. We use gases with temperature $T$ and central chemical potential $\mu$ in the range $65$-$120\,$nK and $k_{\rm B}\times (35$-$60)\,$nK, respectively. The interaction energy per particle is given by $U_{\rm int}=(\hbar^2 \tilde{g}/m)n$ \cite{Yefsah:2011}, where $n$ is the 2D spatial density (typically $100\,$atoms$/\mu$m$^{2}$ in the center), $m$ the atomic mass and $\tilde{g}$ the dimensionless interaction strength. Here $\tilde g=\sqrt{8\pi}a/l_z = 0.093$, where $a=5.3\,$nm is the 3D scattering length and $\l_z=\sqrt{\hbar/ m \omega_z}$ \cite{Yefsah:2011}. The energy $\hbar \omega_z$ ($k_{\rm B}\times 70\,$nK) is comparable to $k_{\rm B}T$ and  $U_{\rm int}$ ($\sim k_{\rm B}\times 40\,$nK at the trap center), and the gas is in the quasi-2D regime.

We stir the cloud with a laser beam which creates a repulsive potential with height $V_{\rm stir}\approx k_{\rm B}\times 80\,$nK. This is at least four times the local chemical potential $\mu_{\rm loc}(r)=\mu-V(r)$. The beam has a Gaussian profile with a waist of $w_0 = 2.0\,(5)\,\mu$m, which is larger than the local healing length $\xi=1/\sqrt{\tilde g n}$ ($\approx 0.3\,\mu$m at the trap center), but small compared to the size of the cloud (full width at half maximum $\approx 25\,\mu$m) (see Fig.\,\ref{fig:scheme}). We stir for typically $t_{\rm stir}=0.2\,$s at constant velocity $v$ in a circle of radius $r$ centred on the cloud. The intensity of the stirring beam is ramped on and off in $\approx 5\,$ms without any significant additional heating. Once the stirring beam is switched off, we let the cloud relax for $0.1\,$s and measure the temperature.

For each configuration ($N$, $T$, $r$), we repeat this experiment for various $v$ from 0 to $2\,$mm/s. We find two different regimes for the response and we show an example of each in Fig.\,\ref{fig:heating}. In Fig.\,\ref{fig:heating}{\bf a}, there is a clear threshold behaviour with no discernable dissipation below a critical velocity. In contrast, in Fig.\,\ref{fig:heating}{\bf b}, the temperature increases without a threshold. We identify these behaviours as the superfluid and normal response, respectively. To model these data we choose for a given configuration the fit function
\begin{equation}
T(v)=T_{v=0}+\kappa\cdot t_{\rm stir}\cdot {\rm max}[(v^2-v_c^2),0] ,
\label{eq:fitFunction}
\end{equation}
which describes the heating of a 2D superfluid in the presence of a moving point-like defect \cite{Astrakharchik:2004}. In equation\,(\ref{eq:fitFunction}) the three fit parameters are the temperature at zero velocity $T_{v=0}$, the heating coefficient $\kappa$, and the critical velocity $v_c$. In the normal state, the fit finds $v_c \sim 0$ and the according quadratic heating stems from the linear scaling of the drag force. Scattering of photons from the stirring beam leads to a `background heating' of less than 10\,nK compared to the temperature $T_0$ before stirring. In the following, we use the mean temperature $T=(T_{v=0}+T_0)/2$ to characterize the cloud.

\begin{figure}
	\begin{center}
		\includegraphics[width=1.0\columnwidth]{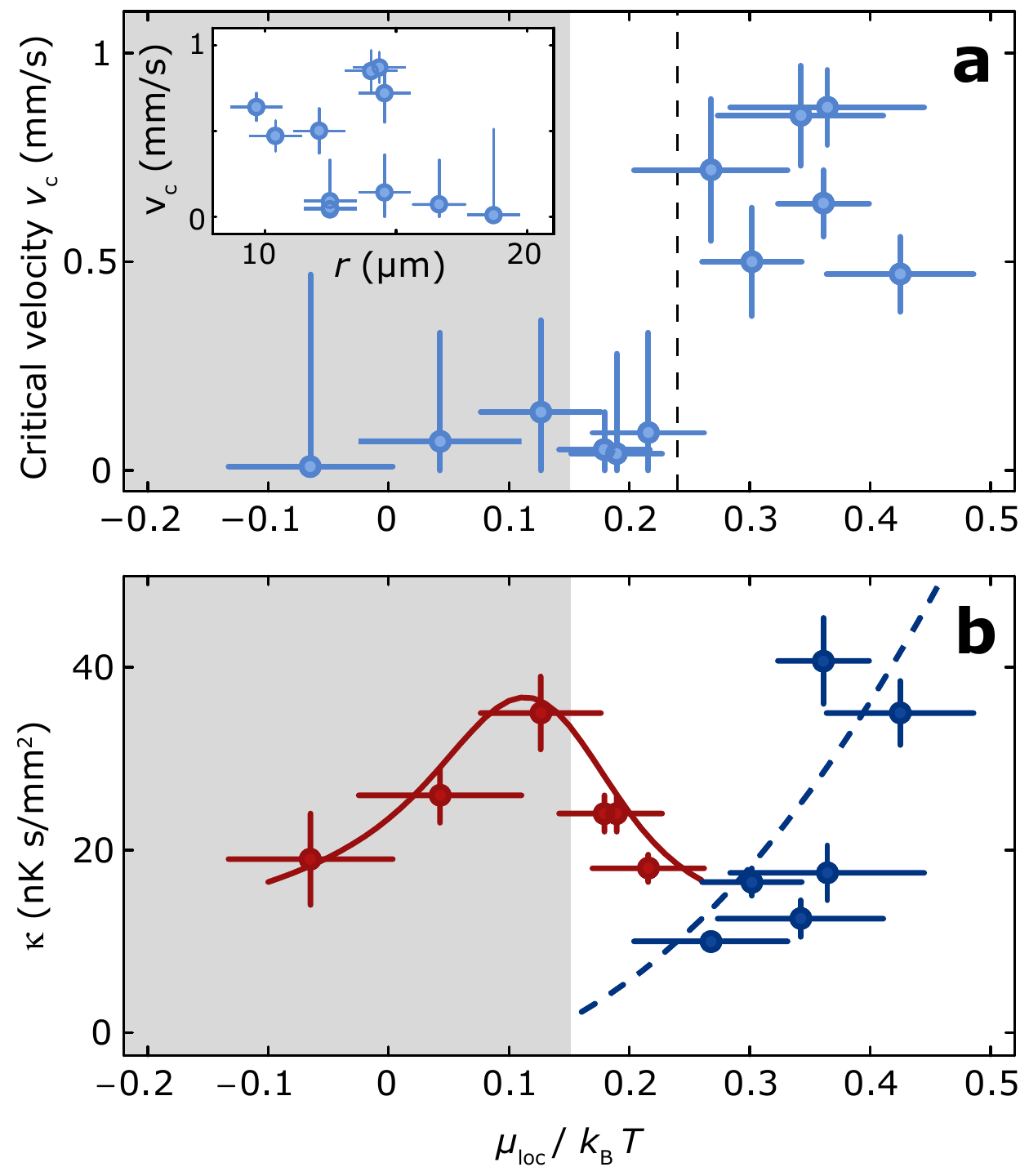}
	\end{center}
\vspace{-0.3cm}
\caption{{\bf Superfluid behaviour across the BKT transition}. {\bf a}, The critical velocities $v_c$ obtained from the curves as in Fig.(\ref{fig:heating}) plotted versus the single parameter $\mu_{\rm loc}/k_{\rm B}T$, which is the relevant quantity due to the scale invariance of the weakly-interacting 2D Bose gas.
Our data show an apparent threshold between critical velocities compatible with zero and clearly non-zero critical velocities. It is located at $\mu_{\rm loc}/k_{\rm B}T\approx0.24$ (dashed line), somewhat above the prediction $\left(\mu_{\rm loc}/k_{\rm B}T\right)_c=0.15$ for the BKT transition in an infinite uniform system \cite{Prokofev:2002} (the grey shaded area indicates the normal state by this prediction).
The $x$ error bars indicate the region of $\mu_{\rm loc}/k_{\rm B}T$ that is traced by the stirring beam due to its size (using the $1/\sqrt{e}$ width of the beam) and due to the `background heating'. The $y$ error bar is the fitting error.
The inset to {\bf a} shows the critical velocity plotted versus the stirring radius $r$. Due to the different atom numbers and temperatures of the clouds, we can find superfluid or normal behaviour for the same radius.
{\bf b}, The heating coefficient $\kappa$ as a function of $\mu_{\rm loc}/k_{\rm B}T$ for the normal data (red circles) and the superfluid data (blue circles).
The red solid line shows a fit of $\kappa$ linear in the normal density, as expected from a single-particle model. The blue dashed line shows an empirical fit quadratic in the superfluid density. The calculation for the densities assumes $T=90\,$nK and the densities are averaged over the size of the stirring beam.
\label{fig:summary}}\vspace{-0.3cm}
\end{figure}

In Fig.\,\ref{fig:summary}, we summarize our data obtained for different configurations ($N$, $T$, $r$). We show in Fig.\,\ref{fig:summary}{\bf a} the fitted critical velocities versus the single parameter $\mu_{\rm loc}(r)/k_{\rm B}T$. The relevance of this parameter results from (i) the local character of the excitation, so that the response of the fluid to the moving perturbation is expected to be similar to that of a uniform gas with the same temperature and the chemical potential $\mu_{\rm loc}$, (ii) the scale invariance of the weakly-interacting 2D Bose gas, whose thermodynamical properties do not depend separately on $\mu$ and $T$, but only on the ratio $\mu/k_{\rm B}T$ \cite{Prokofev:2002,Hung:2011,Yefsah:2011}. In particular, this ratio is univocally related to the phase space density, and thus characterizes the degree of degeneracy of the cloud.

Quite remarkably, the ensemble of our data when plotted as a function of $\mu_{\rm loc}/k_{\rm B}T$ shows an apparent threshold between values compatible with zero and clearly non-zero values. This threshold is located at $\mu_{\rm loc}/k_{\rm B}T\approx 0.24$, somewhat above the prediction $\left(\mu/k_{\rm B}T\right)_c=0.15$ for the superfluid phase transition in a uniform system \cite{Prokofev:2002} with $\tilde g=0.093$.
The deviation may be attributed to (i) finite-size effects in our trapped atomic clouds; (ii) the non-zero width of the stirring beam, which extends over a significant range of $\mu_{\rm loc}/k_{\rm B}T$ (horizontal error bars in Fig.\,\ref{fig:summary}{\bf a}). The emergence of quasi-long range coherence in a trapped 2D system was found \cite{Tung:2010} to agree with the prediction for a homogeneous system. If the same is true for the superfluid behaviour, this means that effect (ii) dominates and that the stirrer has to stand entirely in the superfluid core in order to lead to a non-zero critical velocity.

We limit the presented stirring radii to $r \geq 10\,\mu$m such that the stirring frequencies $\omega = v / r$ for the relevant velocities $v \sim v_c$ are well below $\omega_{r}$. Indeed, smaller radii correspond to a larger centripetal acceleration. This could lead to additional heating via the phonon analog of synchrotron radiation, as observed in the formally similar context of capillary waves generated by a rotating object \cite{Chepelianskii:2008}.

For a homogeneous system, the value of the critical velocity is limited by two dissipation mechanisms, the excitation of phonons or vortices. For a point-like obstacle \cite{Astrakharchik:2004}, phonon excitation dominates and $v_c$ is equal to the speed of sound, given in the $T=0$ limit by $c_s=\hbar \sqrt{\tilde g n}/m$  ($\approx 1.6\,$mm/s for $n=50\,$atoms$/\mu$m$^{2}$) (this situation is described by the celebrated Landau criterion \cite{Leggett:2006}).
When the obstacle size $w_0$ increases and becomes comparable to $\xi$, dissipation via the nucleation of vortex-antivortex pairs (vortex rings in 3D) becomes significant \cite{Langer:1967,Frisch:1992,Winiecki:1999}. The corresponding $v_c$ is then notably reduced with respect to $c_s$.  In the limit of very large obstacles ($w_0\gg \xi$), an analytical analysis of the superfluid flow stability yields $v_c \sim \hbar/m w_0\ll c_s$  \cite{Stiessberger:2000,Crescimanno:2000}. With an  obstacle size $w_0 \gtrsim \xi$, our experimental situation is  intermediate between these two asymptotic regimes. For a non-homogeneous system like ours with the stirring obstacle close to the border of the expected superfluid regime, one can also excite surface modes \cite{Dalfovo:1997,Dubessy:2012}, which constitute an additional dissipation mechanism.

Our measured critical velocities are in the range $0.5$--$1.0\,$mm/s, \emph{i.e.},  $v_c/c_s = 0.3 - 0.6$. By contrast, previous experiments in 3D clouds found lower fractions $v_c / c_s\sim 0.1$ \cite{Onofrio:2000}. The difference may be due to the larger size of the obstacles that were used, and to the average along the axis of the stirring beam of the density distribution in the 3D gas \cite{Fedichev:2001}. The dominant dissipation mechanism could be revealed by e.g. directly observing the created vortex pairs as in Ref. \cite{Neely:2010} or interferometrically detecting the Cerenkov-like wave pattern for $v>c_s$ as in experiments with a non-equilibrium 2D superfluid of exciton-polariton quasi-particles \cite{Amo:2009}.

Fig.\,\ref{fig:summary}{\bf b} shows the fitted heating coefficients $\kappa$ for the normal (red circles) and superfluid data (blue circles). In the normal region, we expect the heating to scale linearly with the normal density $n_{\rm no}$ \cite{Raman:2001}. Using the prediction of \cite{Prokofev:2002} for $\bar{n}_{\rm no}$ (averaged over the size of the stirring beam) we fit $\kappa=a_1 \cdot \bar{n}_{\rm no}$ and obtain $a_1\approx 3\cdot 10^{-6}\,{\rm nK}\cdot{\rm s}$. This value is in reasonable agreement with the prediction of a model \cite{Raman:2001} of a single particle with a thermal velocity distribution of mean $\bar{v}=\sqrt{\pi k_{\rm B}T/2 m}$ colliding with a moving hard wall of width $L=w_0$ yielding $a_1=16 m L \bar{v} /\pi N k_{\rm B} \sim 6\cdot 10^{-6}\,{\rm nK}\cdot{\rm s}$ (for $N=65000$ and $T=90\,$nK). In particular our data nicely reproduce the maximum of  $\bar{n}_{\rm no}$ around the expected superfluid transition point. In the superfluid case and $v>v_c$, we empirically fit a quadratic scaling of the heating with density $\kappa=a_2 \cdot n_{\rm SF}^2$ and find $a_2 = 8\cdot 10^{-9}\,{\rm nK}\cdot{\rm s}\cdot{\rm \mu m}^2$.
In principle, one could develop a more refined model to describe the superfluid region, by taking into account the coexistence of the normal and superfluid states via the sum of two heating terms. However, within the accuracy of our data, we did not find any evidence for the need of such a more refined description.

In conclusion, we have presented a direct proof of the superfluid character of a trapped 2D Bose gas. We observed friction-less motion of an obstacle below a critical velocity if the local chemical potential and the temperature are such that the ratio $\mu_{\rm loc}/k_{\rm B}T$ is larger than a threshold value. An interesting extension of our work would be the study of superfluidity from the complementary point of view of persistent currents, by adapting  to 2D the pioneering experiments performed in 3D toroidal traps \cite{Ryu:2007,Ramanathan:2011,Moulder:2012}.

We would like to thank I. Carusotto, M. Holzmann, S. Nascimb\`{e}ne, H. Perrin, L. Pitaevskii, and W. Zwerger for helpful discussions. We acknowledge funding by IFRAF, ANR (project BOFL), the Alexander von Humboldt foundation (C.W.) and DGA (L.C.). Laboratoire Kastler Brossel is a research unit of \'Ecole normale sup\'erieure and Universit\'e Pierre et Marie Curie, associated with CNRS.

\bibliographystyle{naturemag}

\begin{thebibliography}{10}
\expandafter\ifx\csname url\endcsname\relax
  \def\url#1{\texttt{#1}}\fi
\expandafter\ifx\csname urlprefix\endcsname\relax\def\urlprefix{URL }\fi
\providecommand{\bibinfo}[2]{#2}
\providecommand{\eprint}[2][]{\url{#2}}

\bibitem{Mermin:1966}
\bibinfo{author}{Mermin, N.~D.} \& \bibinfo{author}{Wagner, H.}
\newblock \bibinfo{title}{{Absence of ferromagnetism or antiferromagnetism in
  one- or two-dimensional isotropic Heisenberg models.}}
\newblock \emph{\bibinfo{journal}{Phys. Rev. Lett.}}
  \textbf{\bibinfo{volume}{17}}, \bibinfo{pages}{1133--1136}
  (\bibinfo{year}{1966}).

\bibitem{Minnhagen:1987}
\bibinfo{author}{Minnhagen, P.}
\newblock \bibinfo{title}{{The two-dimensional Coulomb gas, vortex unbinding,
  and superfluid-superconducting films}}.
\newblock \emph{\bibinfo{journal}{Rev. Mod. Phys.}}
  \textbf{\bibinfo{volume}{59}}, \bibinfo{pages}{1001--1066}
  (\bibinfo{year}{1987}).

\bibitem{Bishop:1978}
\bibinfo{author}{Bishop, D.~J.} \& \bibinfo{author}{Reppy, J.~D.}
\newblock \bibinfo{title}{{Study of the superfluid transition in
  two-dimensional $^4$He films}}.
\newblock \emph{\bibinfo{journal}{Phys. Rev. Lett.}}
  \textbf{\bibinfo{volume}{40}}, \bibinfo{pages}{1727--1730}
  (\bibinfo{year}{1978}).

\bibitem{Hadzibabic:2006}
\bibinfo{author}{Hadzibabic, Z.}, \bibinfo{author}{Kr\"{u}ger, P.},
  \bibinfo{author}{Cheneau, M.}, \bibinfo{author}{Battelier, B.} \&
  \bibinfo{author}{Dalibard, J.}
\newblock \bibinfo{title}{{Berezinskii-Kosterlitz-Thouless crossover in a
  trapped atomic gas}}.
\newblock \emph{\bibinfo{journal}{Nature}} \textbf{\bibinfo{volume}{441}},
  \bibinfo{pages}{1118--1121} (\bibinfo{year}{2006}).

\bibitem{Clade:2009}
\bibinfo{author}{Clad\'{e}, P.}, \bibinfo{author}{Ryu, C.},
  \bibinfo{author}{Ramanathan, A.}, \bibinfo{author}{Helmerson, K.} \&
  \bibinfo{author}{Phillips, W.~D.}
\newblock \bibinfo{title}{{Observation of a 2D Bose gas: From thermal to
  quasicondensate to superfluid}}.
\newblock \emph{\bibinfo{journal}{Phys. Rev. Lett.}}
  \textbf{\bibinfo{volume}{102}}, \bibinfo{pages}{170401}
  (\bibinfo{year}{2009}).

\bibitem{Tung:2010}
\bibinfo{author}{Tung, S.}, \bibinfo{author}{Lamporesi, G.},
  \bibinfo{author}{Lobser, D.}, \bibinfo{author}{Xia, L.} \&
  \bibinfo{author}{Cornell, E.~A.}
\newblock \bibinfo{title}{{Observation of the presuperfluid regime in a
  two-dimensional Bose gas}}.
\newblock \emph{\bibinfo{journal}{Phys. Rev. Lett.}}
  \textbf{\bibinfo{volume}{105}}, \bibinfo{pages}{230408}
  (\bibinfo{year}{2010}).

\bibitem{Leggett:2006}
\bibinfo{author}{Leggett, A.~J.}
\newblock \emph{\bibinfo{title}{{Quantum Liquids: Bose Condensation and Cooper
  Pairing in Condensed-Matter Systems}}} (\bibinfo{publisher}{Oxford Univ.
  Press}, \bibinfo{address}{Oxford}, \bibinfo{year}{2006}).

\bibitem{Raman:1999}
\bibinfo{author}{Raman, C.} \emph{et~al.}
\newblock \bibinfo{title}{{Evidence for a critical velocity in a Bose-Einstein
  condensed gas}}.
\newblock \emph{\bibinfo{journal}{Phys. Rev. Lett.}}
  \textbf{\bibinfo{volume}{83}}, \bibinfo{pages}{2502--2505}
  (\bibinfo{year}{1999}).

\bibitem{Onofrio:2000}
\bibinfo{author}{Onofrio, R.} \emph{et~al.}
\newblock \bibinfo{title}{{Observation of superfluid flow in a Bose-Einstein
  condensed gas}}.
\newblock \emph{\bibinfo{journal}{Phys. Rev. Lett.}}
  \textbf{\bibinfo{volume}{85}}, \bibinfo{pages}{2228--2231}
  (\bibinfo{year}{2000}).

\bibitem{Raman:2001}
\bibinfo{author}{Raman, C.}, \bibinfo{author}{Onofrio, R.},
  \bibinfo{author}{Vogels, J.~M.}, \bibinfo{author}{Abo-Shaeer, J.~R.} \&
  \bibinfo{author}{Ketterle, W.}
\newblock \bibinfo{title}{{Dissipationless flow and superfluidity in gaseous
  Bose-Einstein condensates}}.
\newblock \emph{\bibinfo{journal}{J. Low Temp. Phys.}}
  \textbf{\bibinfo{volume}{122}}, \bibinfo{pages}{99--116}
  (\bibinfo{year}{2001}).

\bibitem{Engels:2007}
\bibinfo{author}{Engels, P.} \& \bibinfo{author}{Atherton, C.}
\newblock \bibinfo{title}{{Stationary and nonstationary fluid flow of a
  Bose-Einstein condensate through a penetrable barrier}}.
\newblock \emph{\bibinfo{journal}{Phys. Rev. Lett.}}
  \textbf{\bibinfo{volume}{99}}, \bibinfo{pages}{160405}
  (\bibinfo{year}{2007}).

\bibitem{Neely:2010}
\bibinfo{author}{Neely, T.~W.}, \bibinfo{author}{Samson, E.~C.},
  \bibinfo{author}{Bradley, A.~S.}, \bibinfo{author}{Davis, M.~J.} \&
  \bibinfo{author}{Anderson, B.~P.}
\newblock \bibinfo{title}{{Observation of vortex dipoles in an oblate
  Bose-Einstein condensate}}.
\newblock \emph{\bibinfo{journal}{Phys. Rev. Lett.}}
  \textbf{\bibinfo{volume}{104}}, \bibinfo{pages}{160401}
  (\bibinfo{year}{2010}).

\bibitem{Miller:2007}
\bibinfo{author}{Miller, D.~E.} \emph{et~al.}
\newblock \bibinfo{title}{{Critical velocity for superfluid flow across the
  BEC-BCS crossover}}.
\newblock \emph{\bibinfo{journal}{Phys. Rev. Lett.}}
  \textbf{\bibinfo{volume}{99}}, \bibinfo{pages}{070402}
  (\bibinfo{year}{2007}).

\bibitem{Yefsah:2011}
\bibinfo{author}{Yefsah, T.}, \bibinfo{author}{Desbuquois, R.},
  \bibinfo{author}{Chomaz, L.}, \bibinfo{author}{G\"{u}nter, K.~J.} \&
  \bibinfo{author}{Dalibard, J.}
\newblock \bibinfo{title}{{Exploring the thermodynamics of a two-dimensional
  Bose gas}}.
\newblock \emph{\bibinfo{journal}{Phys. Rev. Lett.}}
  \textbf{\bibinfo{volume}{107}}, \bibinfo{pages}{130401}
  (\bibinfo{year}{2011}).

\bibitem{Prokofev:2002}
\bibinfo{author}{Prokof'ev, N.} \& \bibinfo{author}{Svistunov, B.}
\newblock \bibinfo{title}{{Two-dimensional weakly interacting Bose gas in the
  fluctuation region}}.
\newblock \emph{\bibinfo{journal}{Phys. Rev. A}} \textbf{\bibinfo{volume}{66}},
  \bibinfo{pages}{043608} (\bibinfo{year}{2002}).

\bibitem{Astrakharchik:2004}
\bibinfo{author}{Astrakharchik, G.~E.} \& \bibinfo{author}{Pitaevskii, L.~P.}
\newblock \bibinfo{title}{{Motion of a heavy impurity through a Bose-Einstein
  condensate}}.
\newblock \emph{\bibinfo{journal}{Phys. Rev. A}} \textbf{\bibinfo{volume}{70}},
  \bibinfo{pages}{013608} (\bibinfo{year}{2004}).

\bibitem{Hung:2011}
\bibinfo{author}{Hung, C.-L.}, \bibinfo{author}{Zhang, X.},
  \bibinfo{author}{Gemelke, N.} \& \bibinfo{author}{Chin, C.}
\newblock \bibinfo{title}{{Observation of scale invariance and universality in
  two-dimensional Bose gases}}.
\newblock \emph{\bibinfo{journal}{Nature}} \textbf{\bibinfo{volume}{470}},
  \bibinfo{pages}{236--239} (\bibinfo{year}{2011}).

\bibitem{Chepelianskii:2008}
\bibinfo{author}{Chepelianskii, A.~D.}, \bibinfo{author}{Chevy, F.} \&
  \bibinfo{author}{Rapha$\rm\ddot{e}$l, E.}
\newblock \bibinfo{title}{{Capillary-gravity waves generated by a slow moving
  object}}.
\newblock \emph{\bibinfo{journal}{Phys. Rev. Lett.}}
  \textbf{\bibinfo{volume}{100}}, \bibinfo{pages}{074504}
  (\bibinfo{year}{2008}).

\bibitem{Langer:1967}
\bibinfo{author}{Langer, J.~S.} \& \bibinfo{author}{Fisher, M.~E.}
\newblock \bibinfo{title}{{Intrinsic critical velocity of a superfluid}}.
\newblock \emph{\bibinfo{journal}{Phys. Rev. Lett.}}
  \textbf{\bibinfo{volume}{19}}, \bibinfo{pages}{560--563}
  (\bibinfo{year}{1967}).

\bibitem{Frisch:1992}
\bibinfo{author}{Frisch, T.}, \bibinfo{author}{Pomeau, Y.} \&
  \bibinfo{author}{Rica, S.}
\newblock \bibinfo{title}{{Transition to dissipation in a model of superflow}}.
\newblock \emph{\bibinfo{journal}{Phys. Rev. Lett.}}
  \textbf{\bibinfo{volume}{69}}, \bibinfo{pages}{1644--1647}
  (\bibinfo{year}{1992}).

\bibitem{Winiecki:1999}
\bibinfo{author}{Winiecki, T.}, \bibinfo{author}{McCann, J.~F.} \&
  \bibinfo{author}{Adams, C.~S.}
\newblock \bibinfo{title}{{Pressure drag in linear and nonlinear quantum
  fluids}}.
\newblock \emph{\bibinfo{journal}{Phys. Rev. Lett.}}
  \textbf{\bibinfo{volume}{82}}, \bibinfo{pages}{5186--5189}
  (\bibinfo{year}{1999}).

\bibitem{Stiessberger:2000}
\bibinfo{author}{Stie{\ss}berger, J.~S.} \& \bibinfo{author}{Zwerger, W.}
\newblock \bibinfo{title}{{Critcal velocity of superfluid flow past large
  obstacles in Bose-Einstein condensates}}.
\newblock \emph{\bibinfo{journal}{Phys. Rev. A}} \textbf{\bibinfo{volume}{62}},
  \bibinfo{pages}{061601(R)} (\bibinfo{year}{2000}).

\bibitem{Crescimanno:2000}
\bibinfo{author}{Crescimanno, M.}, \bibinfo{author}{Koay, C.~G.},
  \bibinfo{author}{Peterson, R.} \& \bibinfo{author}{Walsworth, R.}
\newblock \bibinfo{title}{{Analytical estimate of the critical velocity for
  vortex pair creation in trapped Bose condensates}}.
\newblock \emph{\bibinfo{journal}{Phys. Rev. A}} \textbf{\bibinfo{volume}{62}},
  \bibinfo{pages}{063612} (\bibinfo{year}{2000}).

\bibitem{Dalfovo:1997}
\bibinfo{author}{Dalfovo, F.}, \bibinfo{author}{Giorgini, S.},
  \bibinfo{author}{Guilleumas, M.}, \bibinfo{author}{Pitaevskii, L.} \&
  \bibinfo{author}{Stringari, S.}
\newblock \bibinfo{title}{{Collective and single-particle excitations of a
  trapped Bose gas}}.
\newblock \emph{\bibinfo{journal}{Phys. Rev. A}} \textbf{\bibinfo{volume}{56}},
  \bibinfo{pages}{3840--3845} (\bibinfo{year}{1997}).

\bibitem{Dubessy:2012}
\bibinfo{author}{Dubessy, R.}, \bibinfo{author}{Liennard, T.},
  \bibinfo{author}{Pedri, P.} \& \bibinfo{author}{Perrin, H.}
\newblock \bibinfo{title}{{Critical rotation of an annular superfluid Bose
  gas}}.
\newblock \emph{\bibinfo{journal}{arXiv:1204.6183v1}}  (\bibinfo{year}{2012}).

\bibitem{Fedichev:2001}
\bibinfo{author}{Fedichev, P.~O.} \& \bibinfo{author}{Shlyapnikov, G.~V.}
\newblock \bibinfo{title}{{Critical velocity in cylindrical Bose-Einstein
  condensates}}.
\newblock \emph{\bibinfo{journal}{Phys. Rev. A}} \textbf{\bibinfo{volume}{63}},
  \bibinfo{pages}{045601} (\bibinfo{year}{2001}).

\bibitem{Amo:2009}
\bibinfo{author}{Amo, A.} \emph{et~al.}
\newblock \bibinfo{title}{{Superfluidity of polaritons in semiconductor
  microcavities}}.
\newblock \emph{\bibinfo{journal}{Nature Phys.}} \textbf{\bibinfo{volume}{5}},
  \bibinfo{pages}{805--810} (\bibinfo{year}{2009}).

\bibitem{Ryu:2007}
\bibinfo{author}{Ryu, C.} \emph{et~al.}
\newblock \bibinfo{title}{{Observation of persistent flow of a Bose-Einstein
  condensate in a toroidal trap}}.
\newblock \emph{\bibinfo{journal}{Phys. Rev. Lett.}}
  \textbf{\bibinfo{volume}{99}}, \bibinfo{pages}{260401}
  (\bibinfo{year}{2007}).

\bibitem{Ramanathan:2011}
\bibinfo{author}{Ramanathan, A.} \emph{et~al.}
\newblock \bibinfo{title}{{Superflow in a toroidal Bose-Einstein condensate: An
  atom circuit with a tunable weak link}}.
\newblock \emph{\bibinfo{journal}{Phys. Rev. Lett.}}
  \textbf{\bibinfo{volume}{106}}, \bibinfo{pages}{130401}
  (\bibinfo{year}{2011}).

\bibitem{Moulder:2012}
\bibinfo{author}{Moulder, S.}, \bibinfo{author}{Beattie, S.},
  \bibinfo{author}{Smith, R.~P.}, \bibinfo{author}{Tammuz, N.} \&
  \bibinfo{author}{Hadzibabic, Z.}
\newblock \bibinfo{title}{{Quantised superflow glitches in an annular
  Bose-Einstein condensate}}.
\newblock \emph{\bibinfo{journal}{arXiv:1112.0334v2}}  (\bibinfo{year}{2012}).

\end{thebibliography}

\end{document}